\documentclass[aps,pra,twocolumn,superscriptaddress,amsmath,amssymb,showpacs,notitlepage]{revtex4-1}
\pdfoutput=1
\usepackage{graphicx}
\usepackage{dcolumn}
\usepackage{bm}
\usepackage{hyperref}
\usepackage{tikz}
\usepackage{mathtools}
\usepackage{physics}
\usetikzlibrary{matrix,calc}
\usepackage{braket}
\usepackage{hyperref}
\usepackage{stackengine}
\usepackage{array}
\usepackage{xcolor,color}
\usepackage{lipsum}
\usepackage{stackengine}
\usepackage{enumitem}
\usepackage{courier}
\usetikzlibrary{quantikz}
\usepackage[caption=false]{subfig}
\usepackage{algorithm}
\usepackage{algpseudocode}
\usepackage{xcolor}
\definecolor{color1bg}{HTML}{ECD9ED}

\usetikzlibrary{quantikz}
\newcolumntype{P}[1]{>{\centering\arraybackslash}p{#1}}

\usepackage{array}
\usetikzlibrary{arrows}

\tikzset{
  treenode/.style = {align=center, inner sep=0pt, text centered,
    font=\small\rmfamily},
  arn_n/.style = {treenode, circle, blue, fill=white,font=\fontsize{8}{12}\rmfamily, draw=black, text width=1em, thick},
  arn_z/.style = {treenode, circle, draw =orange, fill=white,font=\fontsize{8}{12}\rmfamily, text width=1em, thick},
   arn_x/.style = {treenode, circle, draw = blue, fill=white,font=\fontsize{8}{12}\rmfamily, text width=1em, thick}
}
\newcolumntype{?}{!{\vrule width 1pt}}

\begin{document}


\title{An Improved Design for All-Photonic Quantum Repeaters}

\author{Ashlesha Patil}
\email{ashlesha@arizona.edu}
\affiliation{Wyant College of Optical Sciences, The University of Arizona, 1630 East University Boulevard, Tucson, AZ 85721.}
\author{Saikat Guha}
\affiliation{Wyant College of Optical Sciences, The University of Arizona, 1630 East University Boulevard, Tucson, AZ 85721.}

\begin{abstract}
All-photonic quantum repeaters use multi-qubit photonic graph states, called repeater graph states (RGS), instead of matter-based quantum memories, for protection against predominantly loss errors. The RGS comprises tree-graph-encoded logical qubits for error correction at the repeaters and physical {\em link} qubits to create entanglement between neighboring repeaters. The two methods to generate the RGS are probabilistic stitching---using linear optical Bell state measurements (fusion)---of small entangled states prepared via multiplexed-probabilistic linear optical circuits fed with single photons, and a direct deterministic preparation using a small number of quantum-logic-capable solid-state emitters. The resource overhead due to fusions and the circuit depth of the quantum emitter system both increase with the size of the RGS. Therefore engineering a resource-efficient RGS is crucial. We propose a new RGS design, which achieves a higher entanglement rate for all-photonic quantum repeaters using fewer qubits than the previously known RGS would. We accomplish this by boosting the probability of entangling neighboring repeaters with tree-encoded link qubits. We also propose a new adaptive scheme to perform logical BSM on the link qubits for loss-only errors. The adaptive BSM outperforms the previous schemes for logical BSM on tree codes when the qubit loss probability is uniform. It reduces the number of optical modes required to perform logical BSM on link qubits to improve the entanglement rate further.
\end{abstract}
\maketitle

\section{Introduction}
\label{sec:intro}
Quantum networks share information over quantum channels such as optical channels using shared entangled states. Noise in the optical channel originating from photon loss in the optical fiber, detector dark clicks, detector inefficiency, etc, limits the maximum achievable entanglement rate of direct transmission of entangled photons over the optical channel~\cite{pirandola2009direct,pirandola2017fundamental,takeoka2014fundamental}. Special purpose network nodes, known as quantum repeaters~\cite{briegel1998quantum}, are used in the quantum network to attain a higher rate than the direct transmission. Quantum repeater architectures that differ in encoding of qubit in photons~\cite{pant2017allOptical,rozpkedek2023all,kaur2024resource}, use of quantum memories\cite{dhara2021subexponential,sangouard2011quantum,guha2015rate}, entanglement routing protocols~\cite{jiang2009quantum} are shown to outperform direct transmission.

Quantum repeaters with matter-based architecture are equipped with quantum memories to protect qubits from loss and Pauli errors. However, this architecture requires long memory coherence times, fast entangling gates between any two quantum memories and efficient light-matter interaction to transduce the photonic qubit into a matter-based qubit. All-photonic quantum repeater architecture replaces the quantum memories with a photonic graph state~\cite{azuma2015all,pant2017allOptical,rozpkedek2023all,kaur2024resource}. In this work, we use dual-rail photonic graph state called repeater graph state (RGS) as the resource for entanglement routing. A subset of the RGS qubits acts as quantum error-correcting codes that protect quantum information from loss and Pauli errors. 



While designing an all-optical quantum repeater, one must consider the resource requirements for preparing the RGS. The two leading proposals for generating large graph states such as the RGS use (1) linear optical Bell state measurement or fusion~\cite{pant2017allOptical,azuma2015all,MercedesThesis,browne2005Fusion} and (2) quantum emitters~\cite{lindner2009proposal,li2022photonic,russo2018photonic, kaur2024resource}. Each method has its own set of challenges. The fusion operation is probabilistic and requires multiplexing, making it resource-intensive. On the other hand, quantum emitter systems can deterministically generate large graph states using only a few emitters. However, their long gate times for entangling gates between emitters cause delayed emission of new photons, which in turn results in losses of the already emitted photons of the RGS~\cite{kaur2024resource}. Improvements in gate times and circuit depths are necessary to make quantum emitters a viable option for repeater graph state generation. The performance of both these methods to generate RGS degrades with the size, specifically the number of qubits and edges in the RGS. 

 The RGS consists of logical and physical qubits entangled with each other. \cite{azuma2015all,pant2017allOptical} use the RGS where the logical qubits are entangled to form a complete graph state or a clique. However, a subgraph of the clique can be used in the RGS if half of its qubits are completely connected to at least the remaining half qubits~\cite{tzitrin2018local,russo2018photonic}. The biclique, or complete bipartite graph~\cite{tzitrin2018local,rudolph102,russo2018photonic}, is the subgraph of the clique that satisfies this condition with the fewest edges. This makes the biclique a potentially more resource-efficient option than the clique-based RGS. 

The logical qubits of the RGS are typically encoded in the highly loss-tolerant tree code~\cite{varnava2006loss,azuma2015all,pant2017allOptical} and held at the quantum repeater for error correction. The physical or link qubits fly away on an optical channel to undergo linear optical fusion operation and entangle neighboring cliques. The fusion succeeds with a probability of 50\%. The success probability can be increased beyond the fundamental limit by using ancilla photons and linear optics~\cite{vanLoockBell,grice2011arbitrarily,bartolucci2021creation,bayerbach2023bell,olivo2018ancilla,stanisic2017generating,fldzhyan2021compact}, non-linear interactions~\cite{kim2001quantum,barrett2004deterministic,welte2021nondestructive,kwiat1998embedded}, hyper-entanglement~\cite{schuck2006complete,barbieri2007complete,li2017hyperentangled}. As the photonic qubits experience loss and Pauli errors, encoding link qubits in an error correction code can increase the probability of success of link fusions in the all-photonic architecture. Photonic logical Bell state measurement (BSM) protocols for different error correcting codes have been studied before~\cite{ewert2016ultrafast,ewert2017ultrafast,lee2019fundamental,ralph2005loss,schmidt2019efficiencies}. \cite{hilaire2021error} introduced two measurement strategies to perform photonic logical BSM on tree codes that can correct both loss and Pauli errors. When using these strategies for BSM on tree-encoded link qubits in quantum repeaters, all physical qubits in the tree codes must be sent over the optical channel. This increases the number of optical modes used per link BSM, negatively impacting the entanglement rate.  Additionally, the proposed one-way and two-way quantum repeater protocols in \cite{hilaire2021error} with tree-encoded link qubits fail to beat the entanglement rate using direct transmission due to lack of multiplexing.

\subsection{Main results}
We explore two avenues to improve the all-photonic repeater architecture: (1) engineering a new, resource-efficient RGS and (2) boosting the link BSM success probability to increase the entanglement rate. Our main contributions are:
\begin{itemize}
     \item We have designed an RGS with tree-encoded link qubits and biclique at the core, achieving a higher entanglement rate than~\cite{pant2017allOptical} using fewer qubits.
     \begin{itemize}
         \item We compare the new RGS with tree-encoded link qubits with 52 fewer qubits than an RGS without tree-encoded link qubits. Envelopes of the rate vs. distance curves taken over different values of repeaters used~\cite{pant2017allOptical} for the two RGSs are of the form $e^{-sL}$, where $L$ is the distance between Alice and Bob. The new RGS reduces the exponent $s$ to roughly 1/3rd of the old RGS.
     \end{itemize}
    \item We have devised a new adaptive scheme to perform logical BSM on tree-encoded qubits using linear optical fusion and single-qubit measurements.
    \item Our adaptive scheme succeeds with higher probability and requires fewer optical modes to perform link BSM for quantum repeaters than the schemes in \cite{hilaire2021error} for loss-only noise.
\end{itemize}

The rest of the paper is organized as follows: We discuss the linear optical fusion measurement and logical X and Z measurements on tree codes in Section~\ref{sec:preliminaries}. We introduce the adaptive BSM on tree codes in Section~\ref{sub:logicalBSM}. We then discuss all-photonic quantum repeater protocol with tree-encoded link qubits in Section~\ref{sec:rep} and conclude in Section~\ref{sec:conclusion}. 
\section{Preliminaries}
\label{sec:preliminaries}
This paper uses a dual-rail photonic graph state as the resource for entanglement distribution. A graph state is a highly entangled state described using a graph $G(V,E)$, where the vertices $V$ represent qubits and the edges $E$ represent pairwise controlled-phase (CZ) gates applied on the qubits~\cite{patil2023clifford}. The stabilizers of the graph state identified by $G(V,E)$ are - $X_i\prod_{j\in\mathcal{N}_i}Z_j \forall i\in V$, where $\mathcal{N}_i$ is the set of all neighboring qubits of $i$ in $G(V,E)$. The fundamental operations used are single-qubit and photonic two-qubit joint measurements on the photonic graph state. This section first reviews the linear optical fusion measurement, followed by logical X and Z measurements on tree codes. 
\subsection{Linear optical fusion}
\label{subsec:prelim_fusion}
Linear optical fusion is a probabilistic operation that projects dual-rail photonic qubits onto two orthogonal Bell states that differ in phase~\cite{patil2023clifford}, with a success probability of $p_f$. The fusion failure can be modeled as single qubit measurements on the input qubits. The fusion used in this work projects onto the Bell states $\frac{\ket{00}\pm\ket{11}}{\sqrt{2}}$ when successful and onto $\ket{01}$ or $\ket{10}$ on failure~\cite{patil2023clifford}. In other words, the fusion success jointly measures the stabilizers $XX$ and $ZZ$ (modulo phase, correctable using single-qubit Pauli operations on the unmeasured qubits), and its failure measures the stabilizer $ZZ$ (modulo phase). In this paper, we have set $p_f=1/2$.


Photon loss is the most common source of error in photonic architectures. It translates to qubit loss in the dual-rail encoding. The linear optical fusion can herald qubit loss~\cite{MercedesThesis,patil2023clifford}. If one or both input qubits to fusion are lost, it results in the third outcome - \textit{fusion loss}, when no stabilizers are measured. Assuming the qubits have a loss probability of $\epsilon$, the fusion success, failure and loss probabilities become $(1-\epsilon)^2p_f$ and $(1-\epsilon)^2(1-p_f)$, and $1-(1-\epsilon)^2$, respectively.

\subsection{Tree codes}
\label{subsec:prelim_treecodes}
Tree codes are widely used for all-photonic quantum repeater architectures~\cite{azuma2015all,pant2017allOptical, kaur2024resource, borregaard2020one} due to their high tolerance to photon loss. They can tolerate up to 50\% loss, the highest possible value for loss tolerance due to the no-cloning theorem~\cite{varnava2006loss}.  

We define \textit{level} of a qubit in a tree code as its hop distance from the root qubit as shown in FIG.~\ref{fig:LBSM_schematic}(a). we focus on \textit{regular} tree codes where each tree level has qubits with an equal number of children. A regular tree of depth $l$ can be uniquely identified using the branching vector $b = [b_0,b_1,\dots,b_{l-1}]$, an array of the number of children at each level starting with the root. Let $\mathcal{L}_i$ be the set of qubits on level $i$ of the tree and $N_b$ are the number of qubits in the tree excluding the root.

\begin{figure*}
    \centering
    \resizebox{\textwidth}{!}{\begin{tikzpicture}[-,>=stealth',level/.style={sibling distance = 5cm/#1,
  level distance = 2cm},
  fusion_s/.style={circle,inner sep=0pt,minimum size=.6cm,xscale=2.6,fill=green!30!,draw=green!30!,font=\sffamily\large\bfseries},
  fusion_f/.style={circle,inner sep=0pt,minimum size=.6cm,xscale=2.6,fill=yellow!50!,draw=yellow!50!,font=\sffamily\large\bfseries},
  fusion_l/.style={circle,inner sep=0pt,minimum size=.6cm,xscale=2.6,fill=red!30!,draw=red!30!,font=\sffamily\large\bfseries},
  fusion_l/.style={circle,inner sep=0pt,minimum size=.6cm,xscale=2.6,fill=red!30!,draw=red!30!,font=\sffamily\large\bfseries},
  mx/.style={circle,inner sep=0pt,minimum size=.6cm,xscale=1,fill=orange!30!,draw=orange!30!,font=\sffamily\large\bfseries},,
  mz/.style={circle,inner sep=0pt,minimum size=.6cm,xscale=1,fill=blue!30!,draw=blue!30!,font=\sffamily\large\bfseries}] 
   \node[fusion_s] (u) at (2.7,-2.) {};
  \node[fusion_f] (u) at (-2.25,-2.) {};
   \node[fusion_l] (u) at (-7.3,-2.) {};
   \node[blue] (u) at (4,-2.) {$XX,ZZ$};
    \node[blue] (u) at (0-1.2,-2.) {$ZZ$};
\node [arn_n] {}
    child{ node [arn_n] {} 
            child{ node [arn_n] {$X$} 
            	child{ node [arn_n] {$Z$} edge from parent node[above left]
                         {}} 
							child{ node [arn_n] {$Z$}}
            }
             child{ node [arn_n] {$X$} 
            	child{ node [arn_n] {$Z$} edge from parent node[above left]
                         {}} 
							child{ node [arn_n] {$Z$}}
            }                      
    }
    child{ node [arn_n] {}
            child{ node [arn_n] {} 
							child{ node [arn_n] {}}
							child{ node [arn_n] {}}
            }
            child{ node [arn_n] {}
							child{ node [arn_n] {}}
							child{ node [arn_n] {}}
            }
		}
  child{ node [arn_n] {}
            child{ node [arn_n] {$Z$} 
							child{ node [arn_n] {$X$}}
							child{ node [arn_n] {$X$}}
            }
            child{ node [arn_n] {$Z$}
							child{ node [arn_n] {$X$}}
							child{ node [arn_n] {$X$}}
            }
		}
  child{ node [arn_n] {$Z$}
            child{ node [arn_n] {$X$} 
							child{ node [arn_n] {$Z$}}
							child{ node [arn_n] {$Z$}}
            }
            child{ node [arn_n] {$X$}
							child{ node [arn_n] {$Z$}}
							child{ node [arn_n] {$Z$}}
            }
		}
; 
\begin{scope}[shift = {(0.5, 0)}]
\node [arn_n] {}
    child{ node [arn_n] {} 
            child{ node [arn_n] {$X$} 
            	child{ node [arn_n] {$Z$} edge from parent node[above left]
                         {}} 
							child{ node [arn_n] {$Z$}}
            }
            child{ node [arn_n] {$X$} 
            	child{ node [arn_n] {$Z$} edge from parent node[above left]
                         {}} 
							child{ node [arn_n] {$Z$}}
            }                      
    }
    child{ node [arn_n] {}
            child{ node [arn_n] {} 
							child{ node [arn_n] {}}
							child{ node [arn_n] {}}
            }
            child{ node [arn_n] {}
							child{ node [arn_n] {}}
							child{ node [arn_n] {}}
            }
		}
  child{ node [arn_n] {}
            child{ node [arn_n] {$Z$} 
							child{ node [arn_n] {$X$}}
							child{ node [arn_n] {$X$}}
            }
            child{ node [arn_n] {$Z$}
							child{ node [arn_n] {$X$}}
							child{ node [arn_n] {$X$}}
            }
		}
  child{ node [arn_n] {$Z$}
            child{ node [arn_n] {$X$} 
							child{ node [arn_n] {$Z$}}
							child{ node [arn_n] {$Z$}}
            }
            child{ node [arn_n] {$X$}
							child{ node [arn_n] {$Z$}}
							child{ node [arn_n] {$Z$}}
            }
		}
; 
\end{scope}

\filldraw[color=blue, fill=blue, very thick](-19.5,-3) circle (.25);
\node[font=\Large] (u) at (-18.25,-3) {=};
\node[font=\Large\rmfamily] (u) at (-20,0) {(a)};
\node[font=\Large\rmfamily] (u) at (-10,0) {(b)};

\node[font=\large\rmfamily] (u) at (-13.65,-1.9) {Level 0};
\node[font=\large\rmfamily] (u) at (-11.5,-3.75) {Level 1};
\node[font=\large\rmfamily] (u) at (-10.6,-5.75) {Level 2};

\begin{scope}[shift = {(-15, 0)}]
\node [arn_n] {$X$}
    child{ node [arn_n] {$X$}
            child{ node [arn_n] {} 
							child{ node [arn_n] {}}
							child{ node [arn_n] {}}
            }
            child{ node [arn_n] {} 
							child{ node [arn_n] {}}
							child{ node [arn_n] {}}
            }
            child{ node [arn_n] {}
							child{ node [arn_n] {}}
							child{ node [arn_n] {}}
            }
		}
; 

\end{scope}
\end{tikzpicture}
}
    \caption{Tree code (a) Encoding a qubit in a [3,2] tree code with depth two (b) The adaptive BSM scheme. The red, yellow, and green ellipses denote fusion loss, failure, and success. The single qubit measurement basis are written inside the circles.}
    \label{fig:LBSM_schematic}
\end{figure*}

\subsubsection{Logical single-qubit measurements}
\label{subsub:logicalSingle}
A qubit is encoded in a tree code by attaching it to the root of the tree cluster state using the CZ operation followed by Pauli X measurements on the qubit and the root. If $\mathcal{L}_1$ is the set of level 1 qubits, the logical operators of the resulting tree-encoded qubit are~\cite{hilaire2021error}- 
\begin{align}
    X_L &= X_i\prod_{j\in \mathcal{C}(i)}Z_j, \quad\quad  i \in \mathcal{L}_1, {\text{and}}
    \label{eq:XL}
\end{align}
\begin{align}
    Z_L &= \prod_{k\in \mathcal{L}_1}Z_i.
    \label{eq:ZL}
\end{align}
Here, $\mathcal{C}(i)$ is the set of children of qubit $i$. From Eq.(~\ref{eq:ZL}), logical Z measurement on the tree-encoded qubit is implemented by performing single-qubit Z measurements on all $\mathcal{L}_1$ qubits. The single qubit Z measurement on a physical qubit is called the \textit{direct}-Z measurement. 

Now we calculate the success probability of the logical measurements assuming all qubits at level-$k$ have $\epsilon_k$ probability of loss.  If a qubit $i$ in the tree code is lost, its Z-measurement ($Z_i$) result can be recovered by performing an \textit{indirect}-Z measurement~\cite{varnava2006loss}. Indirect-Z measurement utilizes the unity eigenvector property of the stabilizers of the state. To perform an indirect-Z measurement on a qubit in $\mathcal{L}_k$, one of its children and the corresponding grandchildren are measured in $X$ and $Z$ bases, respectively. This gives $b_k$ possible attempts for indirect-Z measurement, and at least one of them must succeed. If $k\leq (l-1)$, the success probability indirect-Z measurementis~\cite{varnava2006loss} - 
\begin{align*}
   \xi_k = 1-[1-(1-\epsilon_{k+1})(1-\epsilon_{k+2}+\epsilon_{k+2}\xi_{k+2})^{b_{k+1}}]^{b_k}
\end{align*}
$\xi_{l} = 0, b_{l}=0$. Note that performing $Z$ measurement on a qubit of the tree code refers to performing both direct and indirect measurements simultaneously. The success probability of a Z-measurement is then the weighted sum of the direct and indirect measurement probabilities: $P_{Z,k} = 1-\epsilon_k+\epsilon_k\xi_k$.

The success probability of the logical-Z measurement on the tree code is~\cite{azuma2015all,pant2017allOptical}
\begin{align}
    P_{Z_L} = P_{Z,1}^{b_0}
\end{align}

Similarly, from Eq.(~\ref{eq:XL}), logical X measurement can be performed by measuring any one of the $\mathcal{L}_1$ qubits in the X-basis and all of its children in the Z-basis. Note that this measurement sequence is equivalent to performing an indirect Z measurement on the root. The success probability of $X_L$ measurement is~\cite{azuma2015all,pant2017allOptical}
\begin{align}
    P_{X_L} &= \xi_{0}
\end{align}

A tree code's loss tolerance depends on the number of levels and qubit distribution among the levels, not just the number of qubits. For example, the optimal depth of the tree for the $Z_L$ measurement is two. A depth-two tree can increase $Z$ measurement success probability on $\mathcal{L}_1$ qubits by enabling indirect-Z measurements, unlike a depth-one tree. Making the tree depth-three increases the number of measurements required for indirect-Z measurement of $\mathcal{L}_1$ qubits, decreasing the success probability. Similarly, Z-measurements on $\mathcal{L}_2$ qubits are required for $X_L$ measurement, making the optimal depth three.
\section{Logical BSM on Tree Codes}
\label{sub:logicalBSM}
  
Given two logical qubits with logical operators $X_L,Z_L$ and $X'_L,Z'_L$, the BSM on them is the joint measurement of the operators $X_{L}X'_{L}$ and $Z_{L}Z'_{L}$~\cite{patil2023clifford}. Assuming the two logical qubits are encoded in identical tree codes, using Eq.(\ref{eq:XL})-(\ref{eq:ZL}), 
\begin{align}
    X_{L}X'_L &=  X_i X_{i'}\prod_{j\in \mathcal{C}(i)}Z_j\prod_{j'\in \mathcal{C}(i')}Z_{j'}  \quad i \in \mathcal{L}_1,i' \in\mathcal{L}_{1'}
    \label{eq:BSMXL}
\end{align}
\begin{equation}
    \begin{split}
    Z_{L}Z'_L &= \prod_{j\in \mathcal{L}_1}{Z_j} \prod_{j'\in \mathcal{L}_{1'}}{Z_{i'}}\\&= Z_i Z_{i'}\prod_{j\in \mathcal{L}_1\setminus\{i\}}{Z_j} \prod_{j'\in \mathcal{L}_{1'}\setminus\{i'\}}
     \label{eq:BSMZL}
\end{split}
\end{equation}
The following measurements implement BSM on the logical qubits - 
\begin{itemize}
    \item To satisfy the requirement for measuring $X_i X_{i'}$ and $Z_i Z_{i'}$ operators simultaneously, a BSM is performed on a pair of level-1 qubits from the two trees, namely $i$ and $i'$. 
    \item The remaining operators in $X_{L}X'_L$ and $Z_{L}Z'_L$ are measured by performing Z-measurements on the children of $i$ and $i'$ and the remaining level-1 qubits of the two trees, respectively
\end{itemize}

As the tree codes considered in this work are made of photonic graph states, the BSM on qubits $i$ and $i'$ is replaced by linear optical fusion, making the logical BSM probabilistic. The probability of success of the logical BSM depends upon the measurement pattern of the physical qubits. Two measurement patterns, static and dynamic, are described in \cite{hilaire2021error} to implement logical BSM using fusions and single-qubit measurements. These patterns can correct both loss and Pauli errors. The static BSM attempts pairwise fusions on all qubits of the trees simultaneously. A fusion success on a pair of level-1 qubits and fusion success or failure on all of their children is necessary to get $X_LX_L'$. A fusion on the level-1 qubits contributes to $Z_LZ_L'$ as long as there is no fusion loss. If a fusion on level-$k$ qubits fails, the protocol implements indirect-Z measurements on them using fusion outcomes of their children and grandchildren. This retrieves the $Z$ operators required for $X_LX_L'$ and $Z_LZ_L'$. The dynamic BSM attempts fusions on only the level-1 qubits simultaneously. From level-2 onwards, it performs fusions only if the fusion on the parent is successful; otherwise, it performs single qubit measurements to implement an indirect-Z measurement on the parent. The dynamic protocol achieves a higher logical BSM success probability than the static protocol by performing fewer fusions.

\subsection{Adaptive BSM} 

We have designed a measurement protocol that enhances tree code tolerance against loss-only errors by reducing the required fusions, compared to the static and dynamic BSMs. The adaptive BSM performs either fusions or Z-measurements level-1 qubits. Level-2 and onward qubits undergo single-qubit measurements based on the outcome of the fusion on their respective level-1 qubits. The measurement sequence for the adaptive BSM is as follows (see FIG.~\ref{fig:LBSM_schematic} (b))- 
\begin{itemize}
    \item Attempt fusion on a pair of the level-1 qubits.
    \begin{itemize}
        \item If there is fusion loss, perform single-qubit measurements on children to implement an indirect-Z measurement. Attempt fusion on the next pair of the level-1 qubits.
        \item If there is fusion failure, the $ZZ$ operators are still measured on the level-1 qubits, eliminating the need to measure children. Attempt fusion on the next pair of the level-1 qubits.
        \item If there is fusion success, perform Z measurements on all children of the fused qubits and the remaining level-1 qubits that are not fused yet.
    \end{itemize}
\end{itemize}


If $\epsilon_k$ and $\eta_k = 1-\epsilon_k$ are respectively the loss and survival probabilities of qubits on level-$k$ of the tree code with branching vector $b = [b_0,b_1,\dots,b_{l-1}]$, an attempt to measure $X_LX_L'$ by fusing a pair of level-1 qubits and performing Z-measurements on their children succeeds with the probability -
\begin{equation}
    \begin{split}
       P_{X_LX'_L} & = \eta_1^2p_f P^{b_1}_{Z,2}
    \end{split}
\end{equation}
Here $\eta_1^2p_f$ is the fusion success probability on a pair of level-1 qubits. If the first fusion succeeds on $(i+1)$-th qubit out of the $b_0$ level-1 qubits, s.t. $j$ out of the $i$ fusions have fusion loss, and the rest have failed, the logical BSM success probability of the adaptive BSM is 
\begin{equation}
    \begin{split}
         P^{\rm A}_{\rm BSM} &= \sum_{i=0}^{b_0-1}\bigg[\sum_{j=0}^{i}\binom{i}{j}(1-\eta_1^2)^{j}(\eta_1^2)^{i-j}(1-p_f)^{i-j}\xi_{1}^{2j}\bigg]\\
         & \times P_{X_LX'_L}P_{Z,1}^{b_0-i-1}
    \end{split}
    \label{eq:PlinkBSM}
\end{equation}
$\xi_{1}$ and $P_{Z,1}$ are the indirect-Z and Z measurement success probabilities on level-1 qubits.

\begin{figure}
    \centering
    \includegraphics[scale=0.6]{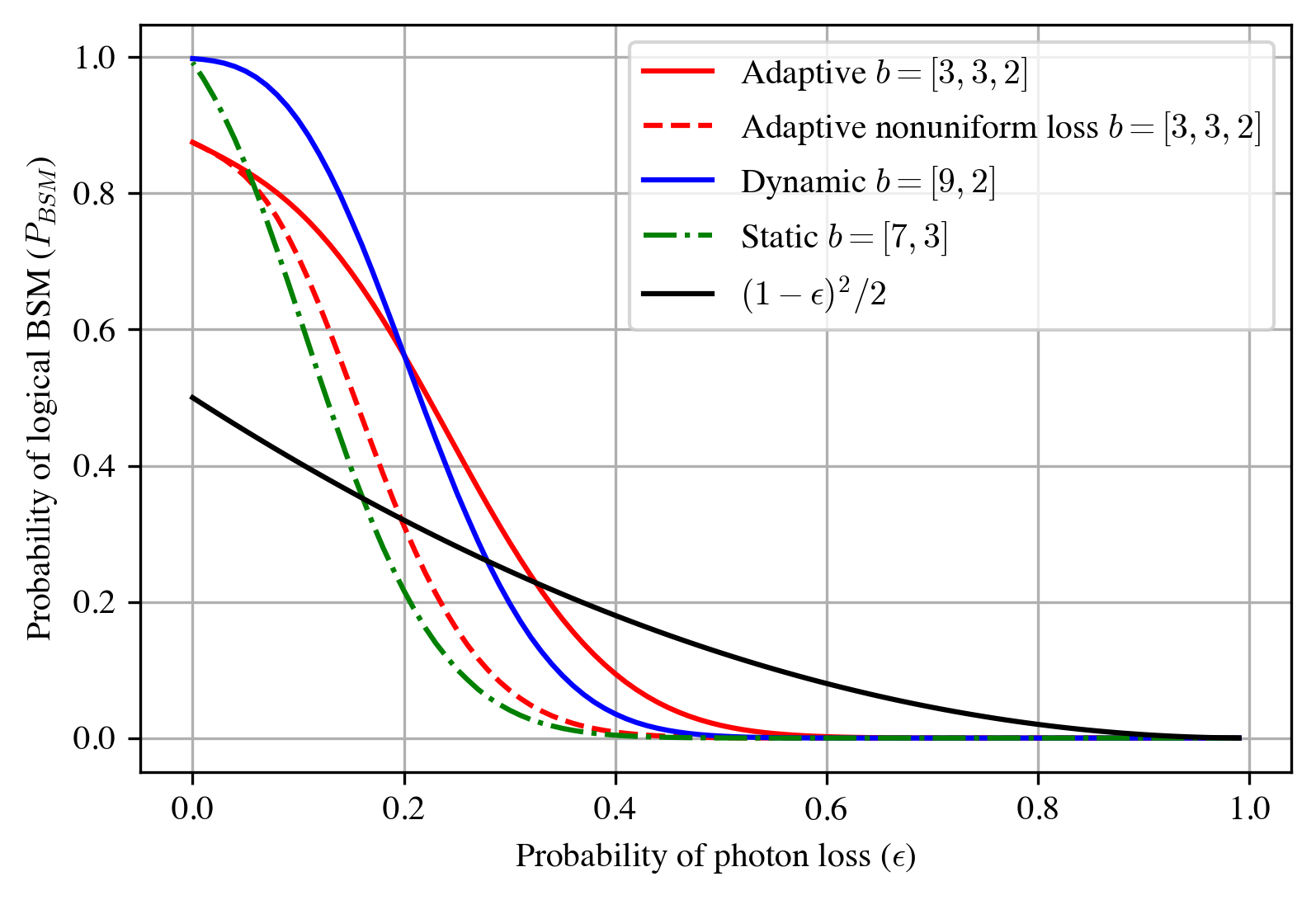}
    \caption{Comparison of logical BSM success probability assuming $p_f=1/2$. The static, dymanic and adaptive BSMs (red solid, green dashed dotted, and blue solid lines, respectively) have uniform loss probability $\epsilon$. For the adaptive nonuniform loss protocol (red dashed line), the loss probabilities are: $\epsilon_1=\epsilon$ and $\epsilon_2=\epsilon_3=1-(1-\epsilon)^2$. The black line signifies fusion on physical qubits.}
    \label{fig:LBSM_comparison}
\end{figure}

The distribution of qubits in the tree code has a greater impact on $P^{\rm A}_{\rm BSM}$ than the number of qubits, similar to logical Pauli measurements on tree codes. The adaptive BSM requires single qubit Z-measurements on the level-2 qubits; from Section~\ref{subsec:prelim_treecodes}, the optimal depth of the tree is three. In low-loss regimes, increasing $b_0$ improves $P^{\rm A}_{\rm BSM}$ by increasing fusion success probability on level-1. However, in high-loss regimes (roughly $\epsilon>0.2$), increasing $b_1$ is better since fusion loss dominates. While increasing $b_0$ leads to more indirect-Z measurements on level-1 qubits, increasing $b_1$ can increase $\xi_1$. FIG.~\ref{fig:LBSM_comparison} shows the logical BSM success probability as a function of loss probability for the static, dynamic, and adaptive BSMs. The regular trees chosen have approximately 30 qubits each and are optimized for the corresponding protocol. The adaptive BSM is better than fusion on physical qubits while $\epsilon<0.33$. It outperforms the dynamic protocol in the high-loss regime.

The choice of the stabilizers measured when fusion succeeds and fails depends on the structure of the logical operators of BSM on the code. The stabilizer operators for logical BSM on tree codes (see Eq.~\ref{eq:BSMXL} and \ref{eq:BSMZL}) have predominantly $Z$ operators. Our selected fusion measures $XX$ and $ZZ$ stabilizers when successful, and $ZZ$ when it fails, to obtain measurement results for the maximum possible number of $Z$ operators in the logical BSM stabilizers. The stabilizers measured by a fusion can be tailored to the logical operators using \textit{rotated} fusions~\cite{patil2023clifford,MercedesThesis,bombin2023increasing}.

In the following section, we use the adaptive BSM to boost the entanglement rate of the all-photonic quantum repeaters.

\section{The All-Photonic Repeater Design}
\label{sec:rep}
The entanglement generation rate of a direct transmission protocol for the pure-loss channel with transmissivity $\eta$ is $R_{\rm direct} = -\log_2(1-\eta)\approx 1.44\eta$~\cite{pirandola2009direct}. This is the \textit{repeaterless rate}. For an optical fiber channel with loss coefficient $\alpha$, the transmissivity decays exponentially with the length of the channel $L$ as $\eta = e^{-\alpha L}$. The repeaterless rate is then $R_{\rm direct} = -\log_2(1-e^{-\alpha L})\approx 1.44e^{-\alpha L}$. Quantum repeaters are placed along the optical channel to surpass $R_{\rm direct}$. ~\cite{pant2017allOptical} gave an all-photonic quantum repeater architecture that beats $R_{\rm direct}$ for pure loss channel. 

In an all-photonic quantum repeater architecture, every repeater holds a dual-rail photonic graph state called the \textit{repeater graph state (RGS)}. The RGS has two types of qubits - link qubits and inner qubits. Inner qubits are logical qubits encoded in the tree code and held at the repeater. The link qubits are used to entangle inner qubits from neighboring repeaters. In~\cite{pant2017allOptical,azuma2015all}, the inner logical qubits form a clique graph state. The complete bipartite, or the \textit{biclique} graph state~\cite{tzitrin2018local,rudolph102,russo2018photonic} can replace the clique graph state without affecting the rate of the protocol in~\cite{tzitrin2018local,russo2018photonic}. Biclique is a special graph state where qubits in one partition are connected to all qubits in the other partition. The RGS with an $N$-qubit biclique has $(N^2-2)/4$ fewer edges than the RGS with an $N$-qubit clique, making the biclique RGS potentially more resource-efficient for state preparation.  

In this paper, we use a biclique RGS, i.e., the inner qubits are entagled to from a biclique as shown in FIG.~\ref{fig:repSchematic}(a). The biclique has $m$ inner qubits in each partition. The qubits of the biclique are encoded in tree code with branching vector $b_{\rm in}$. We assume that all the qubits in the RGS have equal loss probability $\epsilon_{\rm gen}$ after RGS generation. Equivalently, they have survival probability $\eta_{\rm gen} = 1-\epsilon_{\rm gen}$. The assumption holds for RGS created using linear optics~\cite{pant2017allOptical}. If the RGS is generated using quantum emitters, the qubits of the RGS have non-uniform loss probability, and the entanglement generation protocol can be modified accordingly~\cite{kaur2024resource}.

\subsection{The original protocol}
\label{sub:OG}
\begin{figure*}
    \centering
    \includegraphics{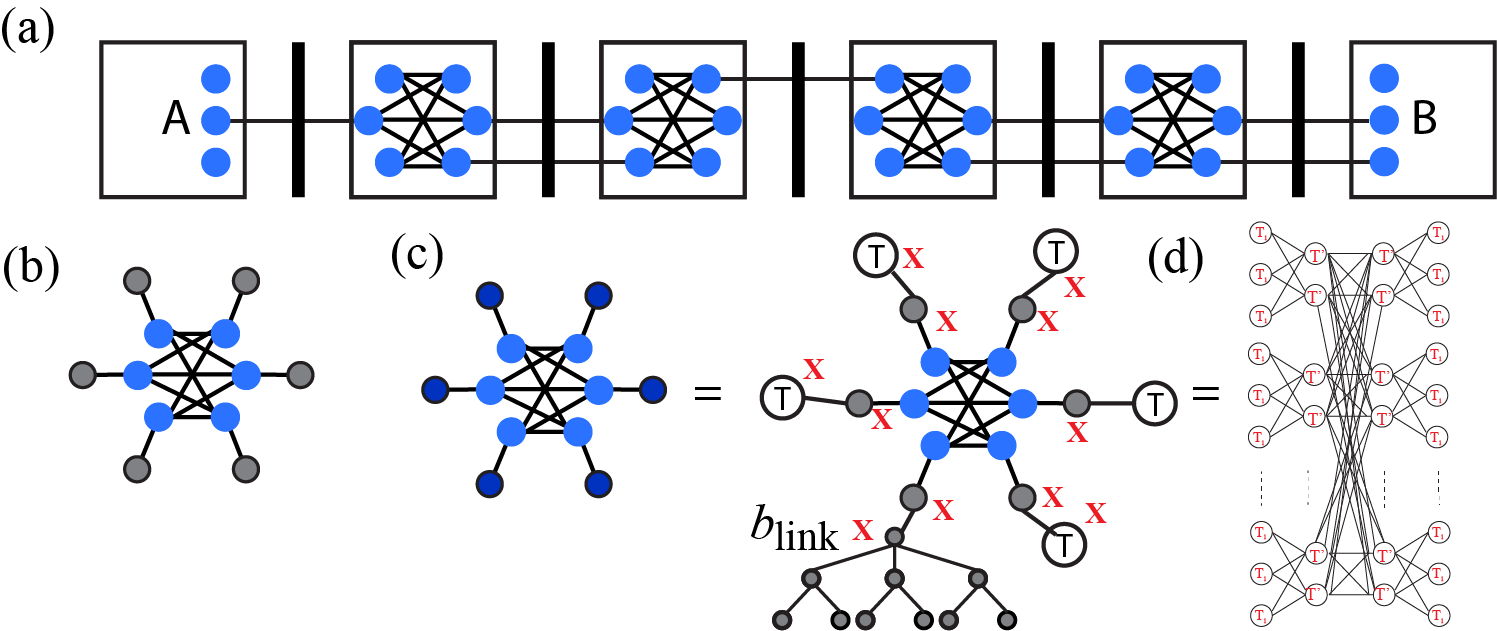}
    \caption{(a) Schematic of a chain of all-photonic quantum quantum repeaters with $n=4$ repeaters and multiplexing $m=3$. The blue circles are tree-encoded logical qubits of the biclique. The branching vector of the tree code is $b_{\rm in} = [b_{i0},b_{i1},\dots,b_{id_i}]$ The black lines connecting the blue circles are the links generated after successful BSM at the minor nodes (solid black rectangles) on the link qubits. (b) The RGS for the original protocol with physical (unencoded) link qubits shown as grey circles and the biclique between tree-encoded logical qubits (c) The RGS for the improved protocol where the link qubits are encoded in tree code with branching vector $b_{\rm link}= [b_{l0},b_{l1},\dots,b_{ld_l}]$. Grey circles are single qubits. (d) The expanded version of the RGS is in (c). Each vertex represents a tree graph state. The edges represent CZ gate between the roots of the trees. The first and fourth columns have $mb_{l0}$ trees with $b'_{T_1}= [b_{l1},\dots,b_{ld_l}]$. Similarly, the second and the third columns have $mb_{i0}$ trees each, representing tree graph state with $b'_{T'}= [b_{i1},\dots,b_{id_i}]$. The roots of the $b_{l0}$ trees in the first (fourth) column form biclique with $b_{i0}$ trees in the second (third) column. Similarly, the roots of the trees in the second and third columns also form a biclique. The figure shown has $b_{l0}=3$ and $b_{i0}=2$. }
    \label{fig:repSchematic}
\end{figure*}

This section reviews the all-photonic repeater protocol from~\cite{pant2017allOptical} but using biclique RGS. The users, Alice and Bob, are $L$ km apart. There are $n$ equidistant repeaters between them. Each repeater has the biclique RGS as shown in FIG.~\ref{fig:repSchematic}(b) such that link qubits are physical qubits and not encoded in the tree code. At the beginning of the entanglement generation protocol, the link photons are sent to the \textit{minor nodes}, placed halfway between neighboring repeaters, to undergo fusion. $m$ fusions coincide at all minor nodes, and each succeeds i.i.d. with probability $p = \eta_{\rm gen}^2\eta^{1/(n+1)} p_{f}$. Here, $\eta^{\frac{1}{2(n+1)}}$ is the survival probability of a link photon through the optical channel to the minor node. The success and failure outcomes of the fusions are communicated back to the neighboring repeaters. Each repeater performs logical X measurements on one inner qubit from each biclique partition with a successful link fusion and logical Z measurements on the remaining $2m-2$ inner qubits. At the end of this step, Alice and Bon can potentially have a shared Bell state. Due to classical communication delays, the qubits held at the repeaters experience more loss than the link qubits. Consequently, all tree codes at each repeater share a uniform loss probability of $\epsilon = 1-\eta_{\rm gen}\eta^{1/(n+1)}$ across all levels.

The entanglement generation rate is given by -
\begin{align}
    R_{O}&=\frac{P_{X_L}^{2n}P_{Z_L}^{2{(m-1)n}}[1-(1-p)^m]^{(n+1)}}{2m} \textrm{ebits/mode}
    \label{eq:rate}
\end{align}
Here, $P_{X_L}$ and $P_{Z_L}$ are the probabilities of X- and Z- measurements on the tree-encoded qubits of the biclique, respectively, and are functions of $\epsilon$ and $b_{\rm in}$ (see Eq.~\ref{eq:XL}-\ref{eq:ZL}). The rate is inversely proportional to $2m$, the number of optical modes used to send $m$ dual-rail link qubits.

\subsection{Improved protocol with tree-encoded link qubits}
\label{subsec:improved}

The entanglement generation rate is directly proportional to the success probability of the link BSM. In the improved protocol, apart from the inner qubits, we also encode the link qubits in tree code with branching vector $b_{\rm link} = [b_{l0},b_{l1},\dots]$ to boost the link BSM success probability. 

FIG.~\ref{fig:repSchematic}(c)-(d) shows the RGS for the improved protocol. The BSMs are performed on the link tree codes using our adaptive protocol from Section~\ref{sub:logicalBSM}. In this protocol, the repeaters send only the $b_{l0}$ level-1 qubits of each link tree code to minor nodes to undergo fusion while the remaining qubits stay at the repeaters. 
The outcomes of all level-1 fusions are then communicated back to the neighboring repeaters. These outcomes are used to perform single qubit measurements on level 2 and onward qubits of the tree $b_{\rm link}$ at the repeater to complete the logical BSM. Based on which link BSM succeeds, the repeater performs logical X and Z measurements on the inner qubits. 

For the link tree code, the survival probability of the level-1 qubits is $\eta_1 = \eta_{\rm gen}\eta^{\frac{1}{2(n+1)}}$. Assuming that the photonic measurements are instantaneous, all the qubits at the repeaters, including the level-2 onwards qubits of the link tree codes, have loss probability $\epsilon = 1-\eta_{\rm gen}\eta^{1/(n+1)}$. The new link BSM success probability $P^{\rm A}_{\rm BSM}$, is calculated using Eq.(~\ref{eq:PlinkBSM}) The rate becomes - 

\begin{align}
    R_{I,A}&=\frac{P_{X_L}^{2n}P_{Z_L}^{2{(m-1)n}}[1-(1-P^{\rm A}_{\rm BSM})^m]^{(n+1)}}{2mb_{l0}} \textrm{ebits/mode}
    \label{eq:rate}
\end{align}
 $P_{X_L}$ and $P_{Z_L}$ are the probabilities of X and Z measurements on inner qubits of the biclique encoded in tree $b_{\rm in}$ (see Eq.~\ref{eq:XL}-\ref{eq:ZL}). $P^{\rm A}_{\rm BSM}$ the adaptive BSM success probability on the tree $b_{\rm link}$. In the improved protocol with adaptive BSM, $2mb_{l0}$ optical modes are used to send the level-1 link qubits to minor nodes.

If the static or the dynamic BSM is used for the link BSM instead of the adaptive BSM, the repeater sends the entire $b_{\rm link}$ tree to the minor nodes. In this case, the $b_{\rm link}$ tree has uniform loss probability $\epsilon = 1-\eta_{\rm gen}\eta^{\frac{1}{2(n+1)}}$. The link BSM outcomes, as before, decide the measurement pattern on the inner qubits. The rate equation changes to 
 \begin{align}
    R_{I,S(D)}&=\frac{P_{X_L}^{2n}P_{Z_L}^{2{(m-1)n}}[1-(1-P^{\rm S(D)}_{\rm BSM})^m]^{(n+1)}}{2mN_{b_{\rm link}}} \textrm{ebits/mode}
    \label{eq:rate_sd}
\end{align}
Here, the subscript $S(D)$ denote the static(dynamic) BSM. $P^{\rm S(D)}_{\rm BSM}$ is the logical BSM success probability for the static (dynamic) BSM and $N_{b_{\rm link}}$ are the number of qubits in the tree $b_{\rm link}$ excluding the root.

\subsection{Entanglement rate}
\label{subsec:results}
\begin{figure}
    \centering
    \includegraphics[scale=0.65]{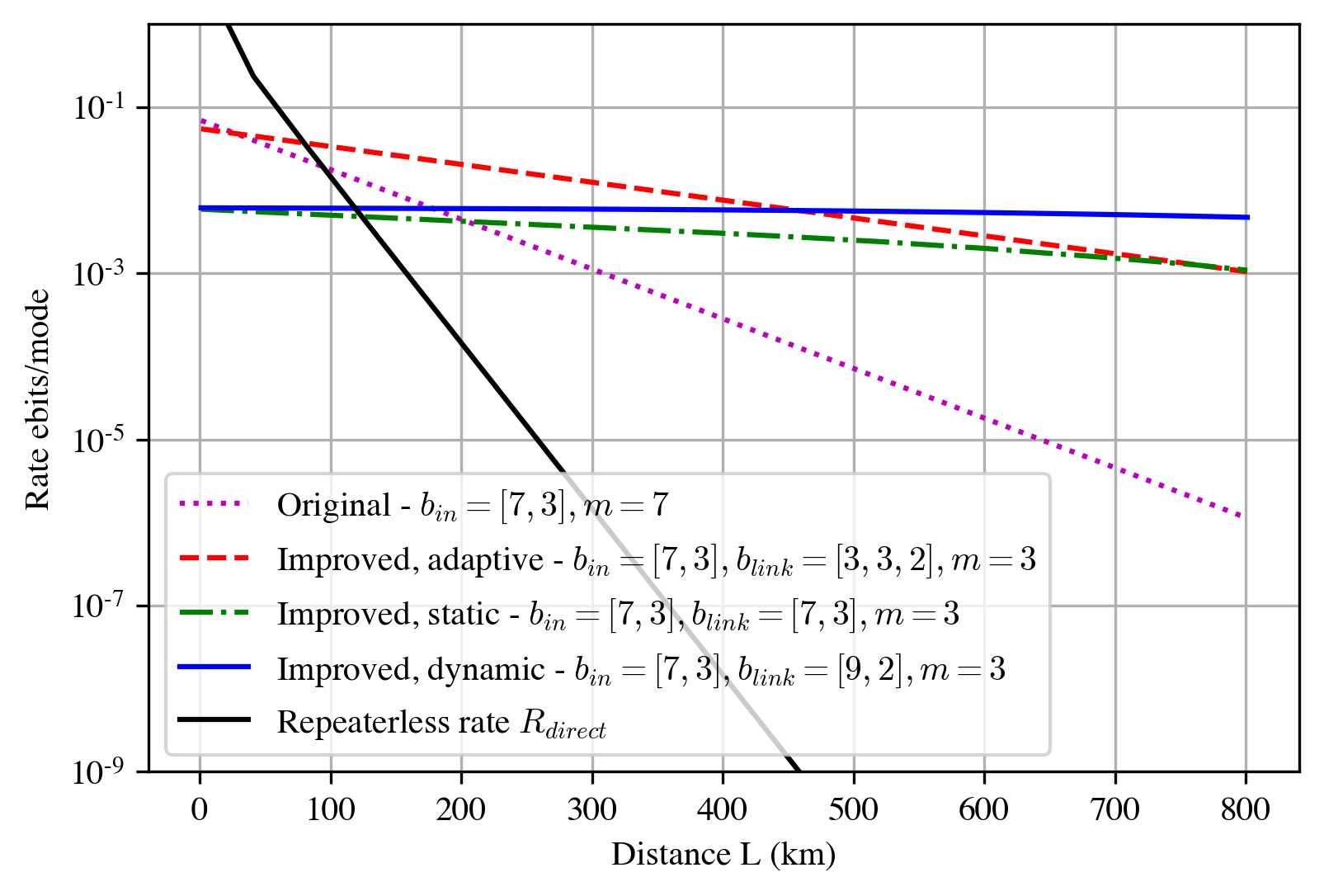}
    \caption{Rate vs. distance envelopes for the original and improved protocols. The number of qubits in the RGSs for the original, the improved with adaptive BSM, static BSM, and dynamic BSM protocols, are 406, 354, 348, and 342, respectively. We assume $\eta_{\rm gen} = 0.9797$ for all RGSs based on \cite{pant2017allOptical}.}
    \label{fig:rateComparison}
\end{figure}

We compare the entanglement rates of the original and the improved protocol for RGSs with roughly the same number of qubits. We also compare the entanglement rates of the different link BSM schemes for the improved protocol. We plot the envelope of entanglement generation rate curves for different values of $n$ vs. distance in FIG.~\ref{fig:rateComparison}. The $b_{\rm in}$ and $b_{\rm link}$ used are optimized for the corresponding protocol. The improved protocol beats the original protocol, irrespective of the type of measurement scheme used for the link BSM. When we fit curves of the form $R=e^{-sL}$ to the rate envelopes, the $s$ of the improved protocol is 1/3rd of the original protocol for the chosen RGSs.

In the improved protocol, the static and dynamic BSMs have uniform loss probability $\epsilon = 1-\eta_{\rm gen}\eta^{\frac{1}{2(n+1)}}$. When using adaptive BSM, the level-1 qubits experience the same loss probability $\epsilon$. However, level-2 onwards qubits wait until the repeater receives the level-1 fusion outcomes to be measured and undergo more loss. The $b_{\rm link}$ tree has a non-uniform loss probability. The adaptive BSM's success probability significantly decreases in this case. We plot the success probability of the adaptive BSM with non-uniform loss probability in FIG.~\ref{fig:LBSM_comparison} assuming $\eta_{\rm gen}=1$.  It still outperforms the static BSM in the high-loss regime. When used for link BSM, the static protocol uses $2N_{b_{\rm link}}$ optical modes per link BSM, compared to $2b_{l0}$ of the adaptive BSM. This results in a higher entanglement rate of the adaptive BSM than the static BSM till roughly 800 km. The success probability of the adaptive BSM with nonuniform loss is strictly worse than the dynamic protocol. The more optical modes the dynamic protocol uses for link BSM compensate for this effect. As a result, there is a regime of $L$ and $n$ where the adaptive BSM outperforms the dynamic BSM.

\section{Conclusion and discussion}
\label{sec:conclusion}

In this work, we have studied a new design for all-optical quantum repeaters for loss-only errors. We have changed the core of the repeater graph state (RGS) from a clique graph state to a biclique graph state. It reduces the number of edges in the RGS, thus potentially reducing its resource requirements. We improve the entanglement rate by encoding the link qubits of the RGS in tree codes and protecting them from losses. We have designed a new measurement scheme for logical BSM on tree codes that uses linear optical fusions and single qubit measurements. This measurement scheme decides the measurement basis on a pair of qubits based on all previous measurement outcomes. It improves upon the previous schemes by reducing the number of probabilistic fusions performed and the number of optical channels used for link BSM. In the improved RGS, we decreased the degree of multiplexing, effectively the size of the biclique, and redirected those extra qubits to boost the link BSM success probability. By strategically redistributing the RGS qubits, we have reduced the size of the RGS and optimized the use of resources while improving the entanglement rate.

In this work, we assumed photon loss as the only source of error. The static and dynamic protocols achieve error correction against depolarization noise~\cite{hilaire2021error}. A natural extension of our work is the performance evaluation of the improved protocol for loss and depolarisation noise. We conjecture that the improved protocol will tolerate both loss and Pauli errors for sufficiently large tree codes. Moreover, comparing our improved RGS with the original RGS with respect to generation complexity using linear optics and quantum emitters would be interesting.

\begin{acknowledgments}
This work was co-funded by the National Science Foundation (NSF) Center for Quantum Networks under grant number 1941583 and NSF grant number 2106189. AP thanks Filip Rozpedek for useful discussions on logical BSM using linear optics.
\end{acknowledgments}

\bibliography{bibFile}

\begin{thebibliography}{43}%
\makeatletter
\providecommand \@ifxundefined [1]{%
 \@ifx{#1\undefined}
}%
\providecommand \@ifnum [1]{%
 \ifnum #1\expandafter \@firstoftwo
 \else \expandafter \@secondoftwo
 \fi
}%
\providecommand \@ifx [1]{%
 \ifx #1\expandafter \@firstoftwo
 \else \expandafter \@secondoftwo
 \fi
}%
\providecommand \natexlab [1]{#1}%
\providecommand \enquote  [1]{``#1''}%
\providecommand \bibnamefont  [1]{#1}%
\providecommand \bibfnamefont [1]{#1}%
\providecommand \citenamefont [1]{#1}%
\providecommand \href@noop [0]{\@secondoftwo}%
\providecommand \href [0]{\begingroup \@sanitize@url \@href}%
\providecommand \@href[1]{\@@startlink{#1}\@@href}%
\providecommand \@@href[1]{\endgroup#1\@@endlink}%
\providecommand \@sanitize@url [0]{\catcode `\\12\catcode `\$12\catcode
  `\&12\catcode `\#12\catcode `\^12\catcode `\_12\catcode `\%12\relax}%
\providecommand \@@startlink[1]{}%
\providecommand \@@endlink[0]{}%
\providecommand \url  [0]{\begingroup\@sanitize@url \@url }%
\providecommand \@url [1]{\endgroup\@href {#1}{\urlprefix }}%
\providecommand \urlprefix  [0]{URL }%
\providecommand \Eprint [0]{\href }%
\providecommand \doibase [0]{http://dx.doi.org/}%
\providecommand \selectlanguage [0]{\@gobble}%
\providecommand \bibinfo  [0]{\@secondoftwo}%
\providecommand \bibfield  [0]{\@secondoftwo}%
\providecommand \translation [1]{[#1]}%
\providecommand \BibitemOpen [0]{}%
\providecommand \bibitemStop [0]{}%
\providecommand \bibitemNoStop [0]{.\EOS\space}%
\providecommand \EOS [0]{\spacefactor3000\relax}%
\providecommand \BibitemShut  [1]{\csname bibitem#1\endcsname}%
\let\auto@bib@innerbib\@empty
\bibitem [{\citenamefont {Pirandola}\ \emph {et~al.}(2009)\citenamefont
  {Pirandola}, \citenamefont {Garc{\'\i}a-Patr{\'o}n}, \citenamefont
  {Braunstein},\ and\ \citenamefont {Lloyd}}]{pirandola2009direct}%
  \BibitemOpen
  \bibfield  {author} {\bibinfo {author} {\bibfnamefont {S.}~\bibnamefont
  {Pirandola}}, \bibinfo {author} {\bibfnamefont {R.}~\bibnamefont
  {Garc{\'\i}a-Patr{\'o}n}}, \bibinfo {author} {\bibfnamefont {S.~L.}\
  \bibnamefont {Braunstein}}, \ and\ \bibinfo {author} {\bibfnamefont
  {S.}~\bibnamefont {Lloyd}},\ }\href@noop {} {\bibfield  {journal} {\bibinfo
  {journal} {Physical review letters}\ }\textbf {\bibinfo {volume} {102}},\
  \bibinfo {pages} {050503} (\bibinfo {year} {2009})}\BibitemShut {NoStop}%
\bibitem [{\citenamefont {Pirandola}\ \emph {et~al.}(2017)\citenamefont
  {Pirandola}, \citenamefont {Laurenza}, \citenamefont {Ottaviani},\ and\
  \citenamefont {Banchi}}]{pirandola2017fundamental}%
  \BibitemOpen
  \bibfield  {author} {\bibinfo {author} {\bibfnamefont {S.}~\bibnamefont
  {Pirandola}}, \bibinfo {author} {\bibfnamefont {R.}~\bibnamefont {Laurenza}},
  \bibinfo {author} {\bibfnamefont {C.}~\bibnamefont {Ottaviani}}, \ and\
  \bibinfo {author} {\bibfnamefont {L.}~\bibnamefont {Banchi}},\ }\href@noop {}
  {\bibfield  {journal} {\bibinfo  {journal} {Nature communications}\ }\textbf
  {\bibinfo {volume} {8}},\ \bibinfo {pages} {1} (\bibinfo {year}
  {2017})}\BibitemShut {NoStop}%
\bibitem [{\citenamefont {Takeoka}\ \emph {et~al.}(2014)\citenamefont
  {Takeoka}, \citenamefont {Guha},\ and\ \citenamefont
  {Wilde}}]{takeoka2014fundamental}%
  \BibitemOpen
  \bibfield  {author} {\bibinfo {author} {\bibfnamefont {M.}~\bibnamefont
  {Takeoka}}, \bibinfo {author} {\bibfnamefont {S.}~\bibnamefont {Guha}}, \
  and\ \bibinfo {author} {\bibfnamefont {M.~M.}\ \bibnamefont {Wilde}},\
  }\href@noop {} {\bibfield  {journal} {\bibinfo  {journal} {Nature
  communications}\ }\textbf {\bibinfo {volume} {5}},\ \bibinfo {pages} {5235}
  (\bibinfo {year} {2014})}\BibitemShut {NoStop}%
\bibitem [{\citenamefont {Briegel}\ \emph {et~al.}(1998)\citenamefont
  {Briegel}, \citenamefont {D{\"u}r}, \citenamefont {Cirac},\ and\
  \citenamefont {Zoller}}]{briegel1998quantum}%
  \BibitemOpen
  \bibfield  {author} {\bibinfo {author} {\bibfnamefont {H.-J.}\ \bibnamefont
  {Briegel}}, \bibinfo {author} {\bibfnamefont {W.}~\bibnamefont {D{\"u}r}},
  \bibinfo {author} {\bibfnamefont {J.~I.}\ \bibnamefont {Cirac}}, \ and\
  \bibinfo {author} {\bibfnamefont {P.}~\bibnamefont {Zoller}},\ }\href@noop {}
  {\bibfield  {journal} {\bibinfo  {journal} {Physical Review Letters}\
  }\textbf {\bibinfo {volume} {81}},\ \bibinfo {pages} {5932} (\bibinfo {year}
  {1998})}\BibitemShut {NoStop}%
\bibitem [{\citenamefont {Pant}\ \emph {et~al.}(2017)\citenamefont {Pant},
  \citenamefont {Krovi}, \citenamefont {Englund},\ and\ \citenamefont
  {Guha}}]{pant2017allOptical}%
  \BibitemOpen
  \bibfield  {author} {\bibinfo {author} {\bibfnamefont {M.}~\bibnamefont
  {Pant}}, \bibinfo {author} {\bibfnamefont {H.}~\bibnamefont {Krovi}},
  \bibinfo {author} {\bibfnamefont {D.}~\bibnamefont {Englund}}, \ and\
  \bibinfo {author} {\bibfnamefont {S.}~\bibnamefont {Guha}},\ }\href@noop {}
  {\bibfield  {journal} {\bibinfo  {journal} {Physical Review A}\ }\textbf
  {\bibinfo {volume} {95}},\ \bibinfo {pages} {012304} (\bibinfo {year}
  {2017})}\BibitemShut {NoStop}%
\bibitem [{\citenamefont {Rozpedek}\ \emph {et~al.}(2023)\citenamefont
  {Rozpedek}, \citenamefont {Seshadreesan}, \citenamefont {Polakos},
  \citenamefont {Jiang},\ and\ \citenamefont {Guha}}]{rozpkedek2023all}%
  \BibitemOpen
  \bibfield  {author} {\bibinfo {author} {\bibfnamefont {F.}~\bibnamefont
  {Rozpedek}}, \bibinfo {author} {\bibfnamefont {K.~P.}\ \bibnamefont
  {Seshadreesan}}, \bibinfo {author} {\bibfnamefont {P.}~\bibnamefont
  {Polakos}}, \bibinfo {author} {\bibfnamefont {L.}~\bibnamefont {Jiang}}, \
  and\ \bibinfo {author} {\bibfnamefont {S.}~\bibnamefont {Guha}},\ }\href@noop
  {} {\bibfield  {journal} {\bibinfo  {journal} {Physical Review Research}\
  }\textbf {\bibinfo {volume} {5}},\ \bibinfo {pages} {043056} (\bibinfo {year}
  {2023})}\BibitemShut {NoStop}%
\bibitem [{\citenamefont {Kaur}\ \emph {et~al.}(2024)\citenamefont {Kaur},
  \citenamefont {Patil},\ and\ \citenamefont {Guha}}]{kaur2024resource}%
  \BibitemOpen
  \bibfield  {author} {\bibinfo {author} {\bibfnamefont {E.}~\bibnamefont
  {Kaur}}, \bibinfo {author} {\bibfnamefont {A.}~\bibnamefont {Patil}}, \ and\
  \bibinfo {author} {\bibfnamefont {S.}~\bibnamefont {Guha}},\ }\href@noop {}
  {\bibfield  {journal} {\bibinfo  {journal} {arXiv preprint arXiv:2402.00731}\
  } (\bibinfo {year} {2024})}\BibitemShut {NoStop}%
\bibitem [{\citenamefont {Dhara}\ \emph {et~al.}(2021)\citenamefont {Dhara},
  \citenamefont {Patil}, \citenamefont {Krovi},\ and\ \citenamefont
  {Guha}}]{dhara2021subexponential}%
  \BibitemOpen
  \bibfield  {author} {\bibinfo {author} {\bibfnamefont {P.}~\bibnamefont
  {Dhara}}, \bibinfo {author} {\bibfnamefont {A.}~\bibnamefont {Patil}},
  \bibinfo {author} {\bibfnamefont {H.}~\bibnamefont {Krovi}}, \ and\ \bibinfo
  {author} {\bibfnamefont {S.}~\bibnamefont {Guha}},\ }\href@noop {} {\bibfield
   {journal} {\bibinfo  {journal} {Physical Review A}\ }\textbf {\bibinfo
  {volume} {104}},\ \bibinfo {pages} {052612} (\bibinfo {year}
  {2021})}\BibitemShut {NoStop}%
\bibitem [{\citenamefont {Sangouard}\ \emph {et~al.}(2011)\citenamefont
  {Sangouard}, \citenamefont {Simon}, \citenamefont {De~Riedmatten},\ and\
  \citenamefont {Gisin}}]{sangouard2011quantum}%
  \BibitemOpen
  \bibfield  {author} {\bibinfo {author} {\bibfnamefont {N.}~\bibnamefont
  {Sangouard}}, \bibinfo {author} {\bibfnamefont {C.}~\bibnamefont {Simon}},
  \bibinfo {author} {\bibfnamefont {H.}~\bibnamefont {De~Riedmatten}}, \ and\
  \bibinfo {author} {\bibfnamefont {N.}~\bibnamefont {Gisin}},\ }\href@noop {}
  {\bibfield  {journal} {\bibinfo  {journal} {Reviews of Modern Physics}\
  }\textbf {\bibinfo {volume} {83}},\ \bibinfo {pages} {33} (\bibinfo {year}
  {2011})}\BibitemShut {NoStop}%
\bibitem [{\citenamefont {Guha}\ \emph {et~al.}(2015)\citenamefont {Guha},
  \citenamefont {Krovi}, \citenamefont {Fuchs}, \citenamefont {Dutton},
  \citenamefont {Slater}, \citenamefont {Simon},\ and\ \citenamefont
  {Tittel}}]{guha2015rate}%
  \BibitemOpen
  \bibfield  {author} {\bibinfo {author} {\bibfnamefont {S.}~\bibnamefont
  {Guha}}, \bibinfo {author} {\bibfnamefont {H.}~\bibnamefont {Krovi}},
  \bibinfo {author} {\bibfnamefont {C.~A.}\ \bibnamefont {Fuchs}}, \bibinfo
  {author} {\bibfnamefont {Z.}~\bibnamefont {Dutton}}, \bibinfo {author}
  {\bibfnamefont {J.~A.}\ \bibnamefont {Slater}}, \bibinfo {author}
  {\bibfnamefont {C.}~\bibnamefont {Simon}}, \ and\ \bibinfo {author}
  {\bibfnamefont {W.}~\bibnamefont {Tittel}},\ }\href@noop {} {\bibfield
  {journal} {\bibinfo  {journal} {Physical Review A}\ }\textbf {\bibinfo
  {volume} {92}},\ \bibinfo {pages} {022357} (\bibinfo {year}
  {2015})}\BibitemShut {NoStop}%
\bibitem [{\citenamefont {Jiang}\ \emph {et~al.}(2009)\citenamefont {Jiang},
  \citenamefont {Taylor}, \citenamefont {Nemoto}, \citenamefont {Munro},
  \citenamefont {Van~Meter},\ and\ \citenamefont {Lukin}}]{jiang2009quantum}%
  \BibitemOpen
  \bibfield  {author} {\bibinfo {author} {\bibfnamefont {L.}~\bibnamefont
  {Jiang}}, \bibinfo {author} {\bibfnamefont {J.~M.}\ \bibnamefont {Taylor}},
  \bibinfo {author} {\bibfnamefont {K.}~\bibnamefont {Nemoto}}, \bibinfo
  {author} {\bibfnamefont {W.~J.}\ \bibnamefont {Munro}}, \bibinfo {author}
  {\bibfnamefont {R.}~\bibnamefont {Van~Meter}}, \ and\ \bibinfo {author}
  {\bibfnamefont {M.~D.}\ \bibnamefont {Lukin}},\ }\href@noop {} {\bibfield
  {journal} {\bibinfo  {journal} {Physical Review A}\ }\textbf {\bibinfo
  {volume} {79}},\ \bibinfo {pages} {032325} (\bibinfo {year}
  {2009})}\BibitemShut {NoStop}%
\bibitem [{\citenamefont {Azuma}\ \emph {et~al.}(2015)\citenamefont {Azuma},
  \citenamefont {Tamaki},\ and\ \citenamefont {Lo}}]{azuma2015all}%
  \BibitemOpen
  \bibfield  {author} {\bibinfo {author} {\bibfnamefont {K.}~\bibnamefont
  {Azuma}}, \bibinfo {author} {\bibfnamefont {K.}~\bibnamefont {Tamaki}}, \
  and\ \bibinfo {author} {\bibfnamefont {H.-K.}\ \bibnamefont {Lo}},\
  }\href@noop {} {\bibfield  {journal} {\bibinfo  {journal} {Nature
  communications}\ }\textbf {\bibinfo {volume} {6}},\ \bibinfo {pages} {6787}
  (\bibinfo {year} {2015})}\BibitemShut {NoStop}%
\bibitem [{\citenamefont {Gimeno-Segovia}(2015)}]{MercedesThesis}%
  \BibitemOpen
  \bibfield  {author} {\bibinfo {author} {\bibfnamefont {M.}~\bibnamefont
  {Gimeno-Segovia}},\ }\emph {\bibinfo {title} {Towards practical linear
  optical quantum computing}},\ \href@noop {} {Ph.D. thesis},\ \bibinfo
  {school} {Imperial College London} (\bibinfo {year} {2015})\BibitemShut
  {NoStop}%
\bibitem [{\citenamefont {Browne}\ and\ \citenamefont
  {Rudolph}(2005)}]{browne2005Fusion}%
  \BibitemOpen
  \bibfield  {author} {\bibinfo {author} {\bibfnamefont {D.~E.}\ \bibnamefont
  {Browne}}\ and\ \bibinfo {author} {\bibfnamefont {T.}~\bibnamefont
  {Rudolph}},\ }\href@noop {} {\bibfield  {journal} {\bibinfo  {journal}
  {Physical Review Letters}\ }\textbf {\bibinfo {volume} {95}},\ \bibinfo
  {pages} {010501} (\bibinfo {year} {2005})}\BibitemShut {NoStop}%
\bibitem [{\citenamefont {Lindner}\ and\ \citenamefont
  {Rudolph}(2009)}]{lindner2009proposal}%
  \BibitemOpen
  \bibfield  {author} {\bibinfo {author} {\bibfnamefont {N.~H.}\ \bibnamefont
  {Lindner}}\ and\ \bibinfo {author} {\bibfnamefont {T.}~\bibnamefont
  {Rudolph}},\ }\href@noop {} {\bibfield  {journal} {\bibinfo  {journal}
  {Physical review letters}\ }\textbf {\bibinfo {volume} {103}},\ \bibinfo
  {pages} {113602} (\bibinfo {year} {2009})}\BibitemShut {NoStop}%
\bibitem [{\citenamefont {Li}\ \emph {et~al.}(2022)\citenamefont {Li},
  \citenamefont {Economou},\ and\ \citenamefont {Barnes}}]{li2022photonic}%
  \BibitemOpen
  \bibfield  {author} {\bibinfo {author} {\bibfnamefont {B.}~\bibnamefont
  {Li}}, \bibinfo {author} {\bibfnamefont {S.~E.}\ \bibnamefont {Economou}}, \
  and\ \bibinfo {author} {\bibfnamefont {E.}~\bibnamefont {Barnes}},\
  }\href@noop {} {\bibfield  {journal} {\bibinfo  {journal} {npj Quantum
  Information}\ }\textbf {\bibinfo {volume} {8}},\ \bibinfo {pages} {11}
  (\bibinfo {year} {2022})}\BibitemShut {NoStop}%
\bibitem [{\citenamefont {Russo}\ \emph {et~al.}(2018)\citenamefont {Russo},
  \citenamefont {Barnes},\ and\ \citenamefont {Economou}}]{russo2018photonic}%
  \BibitemOpen
  \bibfield  {author} {\bibinfo {author} {\bibfnamefont {A.}~\bibnamefont
  {Russo}}, \bibinfo {author} {\bibfnamefont {E.}~\bibnamefont {Barnes}}, \
  and\ \bibinfo {author} {\bibfnamefont {S.~E.}\ \bibnamefont {Economou}},\
  }\href@noop {} {\bibfield  {journal} {\bibinfo  {journal} {Physical Review
  B}\ }\textbf {\bibinfo {volume} {98}},\ \bibinfo {pages} {085303} (\bibinfo
  {year} {2018})}\BibitemShut {NoStop}%
\bibitem [{\citenamefont {Tzitrin}(2018)}]{tzitrin2018local}%
  \BibitemOpen
  \bibfield  {author} {\bibinfo {author} {\bibfnamefont {I.}~\bibnamefont
  {Tzitrin}},\ }\href@noop {} {\bibfield  {journal} {\bibinfo  {journal}
  {Physical Review A}\ }\textbf {\bibinfo {volume} {98}},\ \bibinfo {pages}
  {032305} (\bibinfo {year} {2018})}\BibitemShut {NoStop}%
\bibitem [{\citenamefont {Rudolph}\ and\ \citenamefont
  {Photonics}()}]{rudolph102}%
  \BibitemOpen
  \bibfield  {author} {\bibinfo {author} {\bibfnamefont {T.}~\bibnamefont
  {Rudolph}}\ and\ \bibinfo {author} {\bibfnamefont {A.}~\bibnamefont
  {Photonics}},\ }\href@noop {} {\bibinfo  {journal} {URL http://aip.
  scitation. org/doi/10.1063/1.4976737}\ }\BibitemShut {NoStop}%
\bibitem [{\citenamefont {Varnava}\ \emph {et~al.}(2006)\citenamefont
  {Varnava}, \citenamefont {Browne},\ and\ \citenamefont
  {Rudolph}}]{varnava2006loss}%
  \BibitemOpen
\bibfield  {journal} {  }\bibfield  {author} {\bibinfo {author} {\bibfnamefont
  {M.}~\bibnamefont {Varnava}}, \bibinfo {author} {\bibfnamefont {D.~E.}\
  \bibnamefont {Browne}}, \ and\ \bibinfo {author} {\bibfnamefont
  {T.}~\bibnamefont {Rudolph}},\ }\href@noop {} {\bibfield  {journal} {\bibinfo
   {journal} {Physical review letters}\ }\textbf {\bibinfo {volume} {97}},\
  \bibinfo {pages} {120501} (\bibinfo {year} {2006})}\BibitemShut {NoStop}%
\bibitem [{\citenamefont {Ewert}\ and\ \citenamefont {van
  Loock}(2014)}]{vanLoockBell}%
  \BibitemOpen
  \bibfield  {author} {\bibinfo {author} {\bibfnamefont {F.}~\bibnamefont
  {Ewert}}\ and\ \bibinfo {author} {\bibfnamefont {P.}~\bibnamefont {van
  Loock}},\ }\href@noop {} {\bibfield  {journal} {\bibinfo  {journal} {Physical
  review letters}\ }\textbf {\bibinfo {volume} {113}},\ \bibinfo {pages}
  {140403} (\bibinfo {year} {2014})}\BibitemShut {NoStop}%
\bibitem [{\citenamefont {Grice}(2011)}]{grice2011arbitrarily}%
  \BibitemOpen
  \bibfield  {author} {\bibinfo {author} {\bibfnamefont {W.~P.}\ \bibnamefont
  {Grice}},\ }\href@noop {} {\bibfield  {journal} {\bibinfo  {journal}
  {Physical Review A}\ }\textbf {\bibinfo {volume} {84}},\ \bibinfo {pages}
  {042331} (\bibinfo {year} {2011})}\BibitemShut {NoStop}%
\bibitem [{\citenamefont {Bartolucci}\ \emph {et~al.}(2021)\citenamefont
  {Bartolucci}, \citenamefont {Birchall}, \citenamefont {Gimeno-Segovia},
  \citenamefont {Johnston}, \citenamefont {Kieling}, \citenamefont {Pant},
  \citenamefont {Rudolph}, \citenamefont {Smith}, \citenamefont {Sparrow},\
  and\ \citenamefont {Vidrighin}}]{bartolucci2021creation}%
  \BibitemOpen
  \bibfield  {author} {\bibinfo {author} {\bibfnamefont {S.}~\bibnamefont
  {Bartolucci}}, \bibinfo {author} {\bibfnamefont {P.~M.}\ \bibnamefont
  {Birchall}}, \bibinfo {author} {\bibfnamefont {M.}~\bibnamefont
  {Gimeno-Segovia}}, \bibinfo {author} {\bibfnamefont {E.}~\bibnamefont
  {Johnston}}, \bibinfo {author} {\bibfnamefont {K.}~\bibnamefont {Kieling}},
  \bibinfo {author} {\bibfnamefont {M.}~\bibnamefont {Pant}}, \bibinfo {author}
  {\bibfnamefont {T.}~\bibnamefont {Rudolph}}, \bibinfo {author} {\bibfnamefont
  {J.}~\bibnamefont {Smith}}, \bibinfo {author} {\bibfnamefont
  {C.}~\bibnamefont {Sparrow}}, \ and\ \bibinfo {author} {\bibfnamefont
  {M.~D.}\ \bibnamefont {Vidrighin}},\ }\href@noop {} {\bibfield  {journal}
  {\bibinfo  {journal} {arXiv preprint arXiv:2106.13825}\ } (\bibinfo {year}
  {2021})}\BibitemShut {NoStop}%
\bibitem [{\citenamefont {Bayerbach}\ \emph {et~al.}(2023)\citenamefont
  {Bayerbach}, \citenamefont {D’Aurelio}, \citenamefont {van Loock},\ and\
  \citenamefont {Barz}}]{bayerbach2023bell}%
  \BibitemOpen
  \bibfield  {author} {\bibinfo {author} {\bibfnamefont {M.~J.}\ \bibnamefont
  {Bayerbach}}, \bibinfo {author} {\bibfnamefont {S.~E.}\ \bibnamefont
  {D’Aurelio}}, \bibinfo {author} {\bibfnamefont {P.}~\bibnamefont {van
  Loock}}, \ and\ \bibinfo {author} {\bibfnamefont {S.}~\bibnamefont {Barz}},\
  }\href@noop {} {\bibfield  {journal} {\bibinfo  {journal} {Science Advances}\
  }\textbf {\bibinfo {volume} {9}},\ \bibinfo {pages} {eadf4080} (\bibinfo
  {year} {2023})}\BibitemShut {NoStop}%
\bibitem [{\citenamefont {Olivo}\ and\ \citenamefont
  {Grosshans}(2018)}]{olivo2018ancilla}%
  \BibitemOpen
  \bibfield  {author} {\bibinfo {author} {\bibfnamefont {A.}~\bibnamefont
  {Olivo}}\ and\ \bibinfo {author} {\bibfnamefont {F.}~\bibnamefont
  {Grosshans}},\ }\href@noop {} {\bibfield  {journal} {\bibinfo  {journal}
  {Physical Review A}\ }\textbf {\bibinfo {volume} {98}},\ \bibinfo {pages}
  {042323} (\bibinfo {year} {2018})}\BibitemShut {NoStop}%
\bibitem [{\citenamefont {Stanisic}\ \emph {et~al.}(2017)\citenamefont
  {Stanisic}, \citenamefont {Linden}, \citenamefont {Montanaro},\ and\
  \citenamefont {Turner}}]{stanisic2017generating}%
  \BibitemOpen
  \bibfield  {author} {\bibinfo {author} {\bibfnamefont {S.}~\bibnamefont
  {Stanisic}}, \bibinfo {author} {\bibfnamefont {N.}~\bibnamefont {Linden}},
  \bibinfo {author} {\bibfnamefont {A.}~\bibnamefont {Montanaro}}, \ and\
  \bibinfo {author} {\bibfnamefont {P.~S.}\ \bibnamefont {Turner}},\
  }\href@noop {} {\bibfield  {journal} {\bibinfo  {journal} {Physical Review
  A}\ }\textbf {\bibinfo {volume} {96}},\ \bibinfo {pages} {043861} (\bibinfo
  {year} {2017})}\BibitemShut {NoStop}%
\bibitem [{\citenamefont {Fldzhyan}\ \emph {et~al.}(2021)\citenamefont
  {Fldzhyan}, \citenamefont {Saygin},\ and\ \citenamefont
  {Kulik}}]{fldzhyan2021compact}%
  \BibitemOpen
  \bibfield  {author} {\bibinfo {author} {\bibfnamefont {S.~A.}\ \bibnamefont
  {Fldzhyan}}, \bibinfo {author} {\bibfnamefont {M.~Y.}\ \bibnamefont
  {Saygin}}, \ and\ \bibinfo {author} {\bibfnamefont {S.~P.}\ \bibnamefont
  {Kulik}},\ }\href@noop {} {\bibfield  {journal} {\bibinfo  {journal}
  {Physical Review Research}\ }\textbf {\bibinfo {volume} {3}},\ \bibinfo
  {pages} {043031} (\bibinfo {year} {2021})}\BibitemShut {NoStop}%
\bibitem [{\citenamefont {Kim}\ \emph {et~al.}(2001)\citenamefont {Kim},
  \citenamefont {Kulik},\ and\ \citenamefont {Shih}}]{kim2001quantum}%
  \BibitemOpen
  \bibfield  {author} {\bibinfo {author} {\bibfnamefont {Y.-H.}\ \bibnamefont
  {Kim}}, \bibinfo {author} {\bibfnamefont {S.~P.}\ \bibnamefont {Kulik}}, \
  and\ \bibinfo {author} {\bibfnamefont {Y.}~\bibnamefont {Shih}},\ }\href@noop
  {} {\bibfield  {journal} {\bibinfo  {journal} {Physical Review Letters}\
  }\textbf {\bibinfo {volume} {86}},\ \bibinfo {pages} {1370} (\bibinfo {year}
  {2001})}\BibitemShut {NoStop}%
\bibitem [{\citenamefont {Barrett}\ \emph {et~al.}(2004)\citenamefont
  {Barrett}, \citenamefont {Chiaverini}, \citenamefont {Schaetz}, \citenamefont
  {Britton}, \citenamefont {Itano}, \citenamefont {Jost}, \citenamefont
  {Knill}, \citenamefont {Langer}, \citenamefont {Leibfried}, \citenamefont
  {Ozeri} \emph {et~al.}}]{barrett2004deterministic}%
  \BibitemOpen
  \bibfield  {author} {\bibinfo {author} {\bibfnamefont {M.~D.}\ \bibnamefont
  {Barrett}}, \bibinfo {author} {\bibfnamefont {J.}~\bibnamefont {Chiaverini}},
  \bibinfo {author} {\bibfnamefont {T.}~\bibnamefont {Schaetz}}, \bibinfo
  {author} {\bibfnamefont {J.}~\bibnamefont {Britton}}, \bibinfo {author}
  {\bibfnamefont {W.}~\bibnamefont {Itano}}, \bibinfo {author} {\bibfnamefont
  {J.}~\bibnamefont {Jost}}, \bibinfo {author} {\bibfnamefont {E.}~\bibnamefont
  {Knill}}, \bibinfo {author} {\bibfnamefont {C.}~\bibnamefont {Langer}},
  \bibinfo {author} {\bibfnamefont {D.}~\bibnamefont {Leibfried}}, \bibinfo
  {author} {\bibfnamefont {R.}~\bibnamefont {Ozeri}},  \emph {et~al.},\
  }\href@noop {} {\bibfield  {journal} {\bibinfo  {journal} {Nature}\ }\textbf
  {\bibinfo {volume} {429}},\ \bibinfo {pages} {737} (\bibinfo {year}
  {2004})}\BibitemShut {NoStop}%
\bibitem [{\citenamefont {Welte}\ \emph {et~al.}(2021)\citenamefont {Welte},
  \citenamefont {Thomas}, \citenamefont {Hartung}, \citenamefont {Daiss},
  \citenamefont {Langenfeld}, \citenamefont {Morin}, \citenamefont {Rempe},\
  and\ \citenamefont {Distante}}]{welte2021nondestructive}%
  \BibitemOpen
  \bibfield  {author} {\bibinfo {author} {\bibfnamefont {S.}~\bibnamefont
  {Welte}}, \bibinfo {author} {\bibfnamefont {P.}~\bibnamefont {Thomas}},
  \bibinfo {author} {\bibfnamefont {L.}~\bibnamefont {Hartung}}, \bibinfo
  {author} {\bibfnamefont {S.}~\bibnamefont {Daiss}}, \bibinfo {author}
  {\bibfnamefont {S.}~\bibnamefont {Langenfeld}}, \bibinfo {author}
  {\bibfnamefont {O.}~\bibnamefont {Morin}}, \bibinfo {author} {\bibfnamefont
  {G.}~\bibnamefont {Rempe}}, \ and\ \bibinfo {author} {\bibfnamefont
  {E.}~\bibnamefont {Distante}},\ }\href@noop {} {\bibfield  {journal}
  {\bibinfo  {journal} {Nature Photonics}\ }\textbf {\bibinfo {volume} {15}},\
  \bibinfo {pages} {504} (\bibinfo {year} {2021})}\BibitemShut {NoStop}%
\bibitem [{\citenamefont {Kwiat}\ and\ \citenamefont
  {Weinfurter}(1998)}]{kwiat1998embedded}%
  \BibitemOpen
  \bibfield  {author} {\bibinfo {author} {\bibfnamefont {P.~G.}\ \bibnamefont
  {Kwiat}}\ and\ \bibinfo {author} {\bibfnamefont {H.}~\bibnamefont
  {Weinfurter}},\ }\href@noop {} {\bibfield  {journal} {\bibinfo  {journal}
  {Physical Review A}\ }\textbf {\bibinfo {volume} {58}},\ \bibinfo {pages}
  {R2623} (\bibinfo {year} {1998})}\BibitemShut {NoStop}%
\bibitem [{\citenamefont {Schuck}\ \emph {et~al.}(2006)\citenamefont {Schuck},
  \citenamefont {Huber}, \citenamefont {Kurtsiefer},\ and\ \citenamefont
  {Weinfurter}}]{schuck2006complete}%
  \BibitemOpen
  \bibfield  {author} {\bibinfo {author} {\bibfnamefont {C.}~\bibnamefont
  {Schuck}}, \bibinfo {author} {\bibfnamefont {G.}~\bibnamefont {Huber}},
  \bibinfo {author} {\bibfnamefont {C.}~\bibnamefont {Kurtsiefer}}, \ and\
  \bibinfo {author} {\bibfnamefont {H.}~\bibnamefont {Weinfurter}},\
  }\href@noop {} {\bibfield  {journal} {\bibinfo  {journal} {Physical review
  letters}\ }\textbf {\bibinfo {volume} {96}},\ \bibinfo {pages} {190501}
  (\bibinfo {year} {2006})}\BibitemShut {NoStop}%
\bibitem [{\citenamefont {Barbieri}\ \emph {et~al.}(2007)\citenamefont
  {Barbieri}, \citenamefont {Vallone}, \citenamefont {Mataloni},\ and\
  \citenamefont {De~Martini}}]{barbieri2007complete}%
  \BibitemOpen
  \bibfield  {author} {\bibinfo {author} {\bibfnamefont {M.}~\bibnamefont
  {Barbieri}}, \bibinfo {author} {\bibfnamefont {G.}~\bibnamefont {Vallone}},
  \bibinfo {author} {\bibfnamefont {P.}~\bibnamefont {Mataloni}}, \ and\
  \bibinfo {author} {\bibfnamefont {F.}~\bibnamefont {De~Martini}},\
  }\href@noop {} {\bibfield  {journal} {\bibinfo  {journal} {Physical Review
  A}\ }\textbf {\bibinfo {volume} {75}},\ \bibinfo {pages} {042317} (\bibinfo
  {year} {2007})}\BibitemShut {NoStop}%
\bibitem [{\citenamefont {Li}\ and\ \citenamefont
  {Ghose}(2017)}]{li2017hyperentangled}%
  \BibitemOpen
  \bibfield  {author} {\bibinfo {author} {\bibfnamefont {X.-H.}\ \bibnamefont
  {Li}}\ and\ \bibinfo {author} {\bibfnamefont {S.}~\bibnamefont {Ghose}},\
  }\href@noop {} {\bibfield  {journal} {\bibinfo  {journal} {Physical Review
  A}\ }\textbf {\bibinfo {volume} {96}},\ \bibinfo {pages} {020303} (\bibinfo
  {year} {2017})}\BibitemShut {NoStop}%
\bibitem [{\citenamefont {Ewert}\ \emph {et~al.}(2016)\citenamefont {Ewert},
  \citenamefont {Bergmann},\ and\ \citenamefont {van
  Loock}}]{ewert2016ultrafast}%
  \BibitemOpen
  \bibfield  {author} {\bibinfo {author} {\bibfnamefont {F.}~\bibnamefont
  {Ewert}}, \bibinfo {author} {\bibfnamefont {M.}~\bibnamefont {Bergmann}}, \
  and\ \bibinfo {author} {\bibfnamefont {P.}~\bibnamefont {van Loock}},\
  }\href@noop {} {\bibfield  {journal} {\bibinfo  {journal} {Physical review
  letters}\ }\textbf {\bibinfo {volume} {117}},\ \bibinfo {pages} {210501}
  (\bibinfo {year} {2016})}\BibitemShut {NoStop}%
\bibitem [{\citenamefont {Ewert}\ and\ \citenamefont {van
  Loock}(2017)}]{ewert2017ultrafast}%
  \BibitemOpen
  \bibfield  {author} {\bibinfo {author} {\bibfnamefont {F.}~\bibnamefont
  {Ewert}}\ and\ \bibinfo {author} {\bibfnamefont {P.}~\bibnamefont {van
  Loock}},\ }\href@noop {} {\bibfield  {journal} {\bibinfo  {journal} {Physical
  Review A}\ }\textbf {\bibinfo {volume} {95}},\ \bibinfo {pages} {012327}
  (\bibinfo {year} {2017})}\BibitemShut {NoStop}%
\bibitem [{\citenamefont {Lee}\ \emph {et~al.}(2019)\citenamefont {Lee},
  \citenamefont {Ralph},\ and\ \citenamefont {Jeong}}]{lee2019fundamental}%
  \BibitemOpen
  \bibfield  {author} {\bibinfo {author} {\bibfnamefont {S.-W.}\ \bibnamefont
  {Lee}}, \bibinfo {author} {\bibfnamefont {T.~C.}\ \bibnamefont {Ralph}}, \
  and\ \bibinfo {author} {\bibfnamefont {H.}~\bibnamefont {Jeong}},\
  }\href@noop {} {\bibfield  {journal} {\bibinfo  {journal} {Physical Review
  A}\ }\textbf {\bibinfo {volume} {100}},\ \bibinfo {pages} {052303} (\bibinfo
  {year} {2019})}\BibitemShut {NoStop}%
\bibitem [{\citenamefont {Ralph}\ \emph {et~al.}(2005)\citenamefont {Ralph},
  \citenamefont {Hayes},\ and\ \citenamefont {Gilchrist}}]{ralph2005loss}%
  \BibitemOpen
  \bibfield  {author} {\bibinfo {author} {\bibfnamefont {T.~C.}\ \bibnamefont
  {Ralph}}, \bibinfo {author} {\bibfnamefont {A.}~\bibnamefont {Hayes}}, \ and\
  \bibinfo {author} {\bibfnamefont {A.}~\bibnamefont {Gilchrist}},\ }\href@noop
  {} {\bibfield  {journal} {\bibinfo  {journal} {Physical review letters}\
  }\textbf {\bibinfo {volume} {95}},\ \bibinfo {pages} {100501} (\bibinfo
  {year} {2005})}\BibitemShut {NoStop}%
\bibitem [{\citenamefont {Schmidt}\ and\ \citenamefont {van
  Loock}(2019)}]{schmidt2019efficiencies}%
  \BibitemOpen
  \bibfield  {author} {\bibinfo {author} {\bibfnamefont {F.}~\bibnamefont
  {Schmidt}}\ and\ \bibinfo {author} {\bibfnamefont {P.}~\bibnamefont {van
  Loock}},\ }\href@noop {} {\bibfield  {journal} {\bibinfo  {journal} {Physical
  Review A}\ }\textbf {\bibinfo {volume} {99}},\ \bibinfo {pages} {062308}
  (\bibinfo {year} {2019})}\BibitemShut {NoStop}%
\bibitem [{\citenamefont {Hilaire}\ \emph {et~al.}(2021)\citenamefont
  {Hilaire}, \citenamefont {Barnes}, \citenamefont {Economou},\ and\
  \citenamefont {Grosshans}}]{hilaire2021error}%
  \BibitemOpen
  \bibfield  {author} {\bibinfo {author} {\bibfnamefont {P.}~\bibnamefont
  {Hilaire}}, \bibinfo {author} {\bibfnamefont {E.}~\bibnamefont {Barnes}},
  \bibinfo {author} {\bibfnamefont {S.~E.}\ \bibnamefont {Economou}}, \ and\
  \bibinfo {author} {\bibfnamefont {F.}~\bibnamefont {Grosshans}},\ }\href@noop
  {} {\bibfield  {journal} {\bibinfo  {journal} {Physical Review A}\ }\textbf
  {\bibinfo {volume} {104}},\ \bibinfo {pages} {052623} (\bibinfo {year}
  {2021})}\BibitemShut {NoStop}%
\bibitem [{\citenamefont {Patil}\ and\ \citenamefont
  {Guha}(2023)}]{patil2023clifford}%
  \BibitemOpen
  \bibfield  {author} {\bibinfo {author} {\bibfnamefont {A.}~\bibnamefont
  {Patil}}\ and\ \bibinfo {author} {\bibfnamefont {S.}~\bibnamefont {Guha}},\
  }\href@noop {} {\bibfield  {journal} {\bibinfo  {journal} {arXiv preprint
  arXiv:2312.02377}\ } (\bibinfo {year} {2023})}\BibitemShut {NoStop}%
\bibitem [{\citenamefont {Borregaard}\ \emph {et~al.}(2020)\citenamefont
  {Borregaard}, \citenamefont {Pichler}, \citenamefont {Schr{\"o}der},
  \citenamefont {Lukin}, \citenamefont {Lodahl},\ and\ \citenamefont
  {S{\o}rensen}}]{borregaard2020one}%
  \BibitemOpen
  \bibfield  {author} {\bibinfo {author} {\bibfnamefont {J.}~\bibnamefont
  {Borregaard}}, \bibinfo {author} {\bibfnamefont {H.}~\bibnamefont {Pichler}},
  \bibinfo {author} {\bibfnamefont {T.}~\bibnamefont {Schr{\"o}der}}, \bibinfo
  {author} {\bibfnamefont {M.~D.}\ \bibnamefont {Lukin}}, \bibinfo {author}
  {\bibfnamefont {P.}~\bibnamefont {Lodahl}}, \ and\ \bibinfo {author}
  {\bibfnamefont {A.~S.}\ \bibnamefont {S{\o}rensen}},\ }\href@noop {}
  {\bibfield  {journal} {\bibinfo  {journal} {Physical Review X}\ }\textbf
  {\bibinfo {volume} {10}},\ \bibinfo {pages} {021071} (\bibinfo {year}
  {2020})}\BibitemShut {NoStop}%
\bibitem [{\citenamefont {Bomb{\'\i}n}\ \emph {et~al.}(2023)\citenamefont
  {Bomb{\'\i}n}, \citenamefont {Dawson}, \citenamefont {Nickerson},
  \citenamefont {Pant},\ and\ \citenamefont {Sullivan}}]{bombin2023increasing}%
  \BibitemOpen
  \bibfield  {author} {\bibinfo {author} {\bibfnamefont {H.}~\bibnamefont
  {Bomb{\'\i}n}}, \bibinfo {author} {\bibfnamefont {C.}~\bibnamefont {Dawson}},
  \bibinfo {author} {\bibfnamefont {N.}~\bibnamefont {Nickerson}}, \bibinfo
  {author} {\bibfnamefont {M.}~\bibnamefont {Pant}}, \ and\ \bibinfo {author}
  {\bibfnamefont {J.}~\bibnamefont {Sullivan}},\ }\href@noop {} {\bibfield
  {journal} {\bibinfo  {journal} {arXiv preprint arXiv:2303.16122}\ } (\bibinfo
  {year} {2023})}\BibitemShut {NoStop}%
\end{thebibliography}%

\end{document}